\documentclass[prl,reprint]{revtex4-1}

\usepackage{amsmath}
\usepackage{graphicx}
\usepackage{bm}
\usepackage{epstopdf}
\usepackage{orcidlink}
\usepackage{hyperref}
\hypersetup{
    colorlinks,
    linkcolor={blue!80!blue},
    citecolor={blue!80!blue},
    urlcolor={blue!80!blue}
}

\newcommand{\be}{\begin{equation}}
\newcommand{\ee}{\end{equation}}
\newcommand{\bea}{\begin{eqnarray}}
\newcommand{\eea}{\end{eqnarray}}
\newcommand{\ket}[1]{\left\vert #1    \right\rangle }
\newcommand{\bra}[1]{\left\langle   #1  \right\vert}
\newcommand{\ave}[1]{\left\langle #1   \right\rangle }

\newcommand{\x}{^{\dagger}}
\newcommand{\w}{\omega}
\newcommand{\W}{\Omega}

\begin{document}

\title{Dark Polaron Theory for High Intensity Laser Cooling}
\author{Marcel Morillas-Rozas\,\orcidlink{0009-0005-4570-1016}}
\affiliation{Universidad Politécnica de Cartagena member of European University of Technology EUT+, Research Group of Quantum Technologies, Departamento de Física Aplicada y Tecnología Naval, Cartagena E-30202, Spain}

\author{Alberto López-García\,\orcidlink{0009-0001-0023-9850}}
\affiliation{Universidad Politécnica de Cartagena member of European University of Technology EUT+, Research Group of Quantum Technologies, Departamento de Física Aplicada y Tecnología Naval, Cartagena E-30202, Spain}

\author{Enamul Haque\;\orcidlink{0000-0002-5291-7203}}
\affiliation{Universidad Politécnica de Cartagena member of European University of Technology EUT+, Research Group of Quantum Technologies, Departamento de Física Aplicada y Tecnología Naval, Cartagena E-30202, Spain}

\author{Gonzalo Reina Rivero\,\orcidlink{0000-0003-4219-2306}}
\affiliation{Universidad Politécnica de Cartagena member of European University of Technology EUT+, Research Group of Quantum Technologies, Departamento de Física Aplicada y Tecnología Naval, Cartagena E-30202, Spain}

\author{Javier Cerrillo\,\orcidlink{0000-0001-8372-9953}}
\affiliation{Universidad Politécnica de Cartagena member of European University of Technology EUT+, Research Group of Quantum Technologies, Departamento de Física Aplicada y Tecnología Naval, Cartagena E-30202, Spain}
\email{javier.cerrillo@upct.es}

\begin{abstract}
Conventional laser control schemes for cooling and gate operation of trapped ions are limited to the regime of weak laser intensities and small Lamb-Dicke parameters. To overcome this limitation, we present the concept of dark polarons: spatially extended states of pseudospin polarization that are fully decoupled from a lambda laser configuration. In this picture, all high-order Lamb-Dicke terms collapse into a single linear coupling independent of laser intensity. We apply it to definitively elucidate the reasons behind cooling rate limitations observed in recent experimental implementations of electromagnetically induced transparency with high-intensity lasers.
\end{abstract}

\maketitle

Trapped ions feature prominently in the quantum technological arena for their ability to sustain quantum coherence for long times \cite{Cirac1995, Wineland2013, Friis2018}. Nevertheless, their preparation and operation steps are comparatively slow, limiting overall performance. One of these steps is ground-state laser cooling of the vibrational degrees of freedom. The method of choice is sideband cooling \cite{Wineland1975}, which operates in an electronic transition of small linewidth, enabling spectral resolution of motional sidebands. Since cooling depends on spontaneous emission events, its rate is accordingly curbed in this context. Further, the detrimental effect of carrier scattering in final temperatures imposes the use of weak laser intensities \cite{Cirac1992}. Both aspects severely impact maximum achievable cooling rates.

Electromagnetically induced transparency (EIT) \cite{Morigi2000} addresses both limitations: broad transitions may be used and carrier excitations are avoided. Larger cooling rates proportional to the laser intensity may be achieved, as supported by theoretical predictions based in adiabatic elimination \cite{Morigi2003}. Unfortunately, the perturbative parameter used in the derivation is itself also proportional to the laser intensity, so these predictions are only valid in the weak laser intensity limit. Beyond, numeric simulation point to the existence of an upper bound at large laser intensities \cite{Cerrillo2018, Shankar2019}. This upper bound is all-pervasive and even affects more complex proposals that eliminate blue-sideband heating in addition to carrier excitations \cite{Evers2004, Cerrillo2010, Albrecht2011}. Experimentally, the upper bound had remained elusive \cite{Roos2000} due to the low laser intensities involved, in particular to avoid dark-state depletion from unwanted couplings spoiling the lambda configuration. Recently, though, it has been observed in two experiments \cite{Scharnhorst2018, Jordan2019}, opening up the possibility to explore this regime thoroughly.

In order to produce a theory that can properly describe this effect, a deeper understanding of the problem needs to be gained. It is clear that two aspects can affect cooling rates detrimentally. On the one hand, high-order Lamb-Dicke terms are not incorporated into the picture, and their effect is often to reduce or even cancel red-sideband strengths for high-lying Fock states \cite{Wineland1998}. On the other hand, high laser intensities often translate into slow internal (i.~e.~electronic) dynamics in the EIT context, thus rendering the theoretical predictions of the adiabatic elimination approach inaccurate. Some attempts exist to address high-order Lamb-Dicke effects or the infinite intensity limit \cite{Roghani2008,Zhang2021}, but a picture that is able to cover all regimes simultaneously is still missing.

With the goal of describing this newly accessible regime fully, we present a theory for high intensity laser cooling formed on the grounds of a polaron basis involving both the electronic ground states and the motional degrees of freedom of the ion. In this picture, the perturbative parameter is not proportional to the laser intensity and predictions remain valid in the high intensity limit. We provide a rate equation description for the motional degrees of freedom that closely reproduces the dynamics in all intensity regimes. Polaron master equations have been used to describe strong-coupling effects in the context of quantum transport \cite{Jang2008,Jang2009,Nazir2009,Mccutcheon2010,Mccutcheon2011,Jang2011,Xu2011} but, contrary to the spin-boson system usually addressed, in the trapped-ion context the approach is particularly useful given that all higher-order couplings vanish identically in a three-level system. This allows for the definition of a basis of dark polarons, states which are totally decoupled from laser radiation.

We first introduce the concept of dark and bright polarons, then describe the trapped-ion model under consideration and present the polaron transformation that is used for the development of the theory. We analyze the polaron-transformed model and derive a rate equation for the polaron populations that closely reproduces the cooling rate of the full picture. Finally, we check agreement of this equation in the relevant limits of weak and strong laser intensities.

{\em Model and conventional approach} --- EIT cooling is designed for trapped ions featuring three internal electronic levels in $\Lambda$ configuration: a ground state $\ket\downarrow$, a metastable state $\ket\uparrow$ and a dissipative excited state $\ket e$, with spontaneous decay rate $\Gamma$. The ion is trapped in an approximately harmonic potential, of which we consider for simplicity the shallowest motional mode of frequency $\nu$. Two lasers in Raman configuration induce Rabi frequencies $\W_1$ and $\W_2$. The corresponding Hamiltonian in $\hbar$ units is
\be\begin{split}
H&= \nu b^\dagger b + \omega_e \ket{e} \bra{e} + \omega_{\uparrow} \ket{\uparrow} \bra{\uparrow}+ \omega_{\downarrow} \ket{\downarrow} \bra{\downarrow} \\
&+\Omega_1 \left( \ket{e} \bra{\downarrow} + H.c. \right) \cos\left( \omega_1 t- k_1 x - \phi_1\right) \\
&+\Omega_2 \left( \ket{e} \bra{\uparrow} + H.c. \right) \cos\left( \omega_2 t- k_2 x - \phi_2\right)
\end{split}\ee
where $b$ is the annihilation operator of the motional mode, $\omega_{e,\uparrow,\downarrow}$ are the energies of the respective electronic levels, and $\omega_j$, $k_j$ and $\phi_j$ with $j\in\{1,2\}$ are respectively the frequencies, the wavevector projections on the cooling axis $x$ and the initial phases of the respective laser beams. The Raman condition establishes an overall detuning $\Delta=\omega_e-\omega_{\downarrow}-\omega_1=\omega_e-\omega_{\uparrow}-\omega_2$. It is useful to express each wavevector projection in terms of its corresponding Lamb-Dicke parameter following the definition $\eta_j=k_j x_0$, where $x_0=\sqrt{\frac{\hbar}{2 m\nu}}$ is the zero point motion of the oscillator.
The laser frequencies involved are large compared to all other timescales of the system, which justifies the use of a rotating wave approximation. In the interaction picture with respect to the laser frequencies, the rotating Hamiltonian $H'$ can be split into the vibrational term $H_{m}=\nu b^\dagger b$ and the term involving only the electronic degrees of freedom and the laser driving
\be\begin{split}
H_{e}&=\Delta \ket{e} \bra{e} +\\
&\frac{1}{2} \left( \Omega_1 e^{ik_1x}\ket{e} \bra{\downarrow} + \Omega_2 e^{ik_2x} \ket{e} \bra{\uparrow}   + H.c. \right),\label{EIT}
\end{split}\ee
where the phases $\phi_j$ have been absorbed into a redefinition of states $\ket{\uparrow}$ and $\ket{\downarrow}$. 

\begin{figure}
    \includegraphics[width=\columnwidth, trim={0 0 3cm 0},clip]{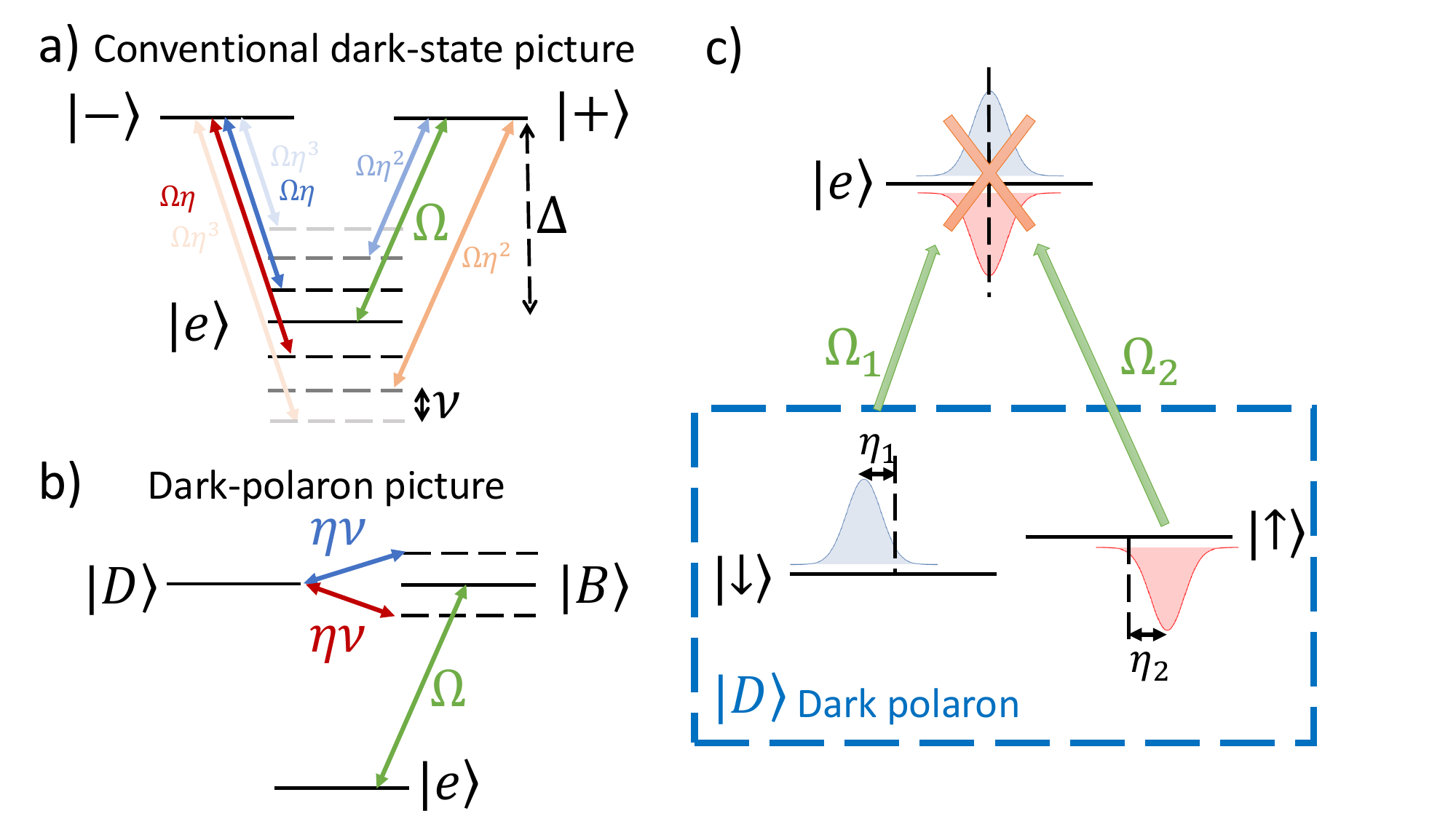}
    \caption{(a) In the conventional EIT picture, laser couplings connect the dark state $\ket -$ to the excited state $\ket e$ through odd-order blue and red sidebands of coupling strengths proportional to $\W\eta^{2n+1}$. Carrier coupling of strength $\W$ and even-order blue and red sidebands of strengths proportional to $\W\eta^{2n}$ connect the bright state $\ket +$ to the excited state $\ket e$. (Not shown: High order effects additionally modulate coupling strengths in terms of Laguerre polynomials \cite{Wineland1998}). (b) Polaron picture of EIT only contains three couplings: a carrier coupling of strength $\Omega$ between the bright polaron $\ket B$ and the excited state and the blue and red sidebands of strength $\eta\nu$ connecting the dark polaron $\ket D$  to the bright polaron. (c) A dark polaron is a superposition of displaced Fock states that cannot be excited by a pair of laser beams.}
    \label{Fig1}
\end{figure}

The conventional EIT picture \cite{Morigi2000} is achieved by expansion in terms of Lamb-Dicke parameters $\eta_j$ and by definition of the dark ($D$) and bright ($B$) basis
\begin{eqnarray}
\ket{D}&=&\frac{1}{\W}\left(\W_1\ket\uparrow-\W_2\ket\downarrow\right),
\label{LDDark}\\
\ket{B}&=&\frac{1}{\W}\left(\W_2\ket\uparrow+\W_1\ket\downarrow\right),
\end{eqnarray}
where $\Omega=\sqrt{\Omega_1^2 + \W_2^2}$. As illustrated in Fig.\ref{Fig1}(a), carrier and even-order sidebands couple the $\ket{B}\leftrightarrow\ket{e}$ transition and odd-order sidebands couple the  $\ket{D}\leftrightarrow\ket{e}$ transition. Usual treatments truncate the expansion to first order, based on the Lamb-Dicke limit condition $\eta_j\sqrt{2\ave{n}+1}\ll1$, with $j\in\{1,2\}$ and $\ave{n}$ the average occupation number of the vibrational mode of the particle. In physical terms, this implies that the recoil energy gained in each photon emission is much smaller than the energy necessary to excite a motional quantum, so processes involving phonon creation or annihilation are realized with small probability. For simplicity, we will consider the case $\eta_1=-\eta_2=\eta$, which is the standard implementation regime of EIT laser cooling. Under these considerations, the Hamiltonian becomes
\be\begin{split}
H_{e}\simeq&\Delta \ket{e} \bra{e} +\frac{\Omega}{2} \left(  \ket{e} \bra{B}  + H.c. \right)\\
&+\frac{\eta\Omega}{2} \left( i \ket{e} \bra{D}  + H.c. \right) \left( b^{\x}  + b\right).\label{LDP}
\end{split}\ee
Transitions towards low Fock states $\ket n$ can be favoured by adjusting the Stark shift of the bright state $\ket{B}$ to $\nu$ with respect to the dark state $\ket{D}$. Ignoring blue sideband effects, adiabatic elimination, valid in the weak coupling limit $\eta\W\ll\nu$, yields a cooling rate prediction
\be
A^-=\frac{\eta^2 \Omega^{2}}{2\Gamma},
\label{eq:Rad}
\ee
This prediction implies that the rate increases together with laser intensity. Nevertheless, as $\W$ increases, both the weak coupling and the Lamb-Dicke limits fail. This rate prediction is invalid for large laser intensity, which calls for a novel theory that decouples $\W$ from the Lamb-Dicke expansion and allows to derive an appropriate rate prediction for increasing laser intensities.

\begin{figure*}
    \includegraphics[width=1.5\columnwidth]{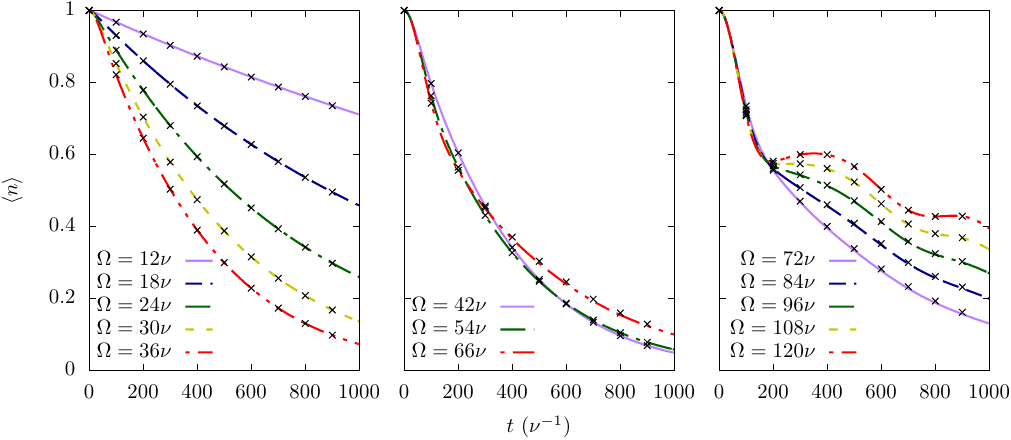}
    \caption{Simulation of laser cooling dynamics for various values of Rabi frequency $\W$ using both the original picture from Eq.(\ref{EIT}) (lines) and the polaron picture from Eqs.(\ref{eq:He},\ref{eq:Hm}) (crosses). Full agreement is reported in all three cooling regimes. On the left, laser cooling rate increases with Rabi frequency. In the center, a cross-over regime appears where the rate stays constant over a large range of Rabi frequencies. On the right, an underdamped regime arises, with  pronounced oscillations and decreasing cooling rate as the Rabi frequency increases.  Parameters: $\Gamma=10\nu$, $\eta_1=-\eta_2=1/\sqrt{2}$, Fock number truncation $n_{max}=20$.}
    \label{Fig2}
\end{figure*}

{\em Polaron picture of the dynamics} ---A tailored polaron transformation yields a picture where a more appropriate perturbative parameter emerges. The transformation, defined by $\tilde{O}=P O P^{\x} $ with
\be
P=\exp\left[ i\left( k_1 \ket{\downarrow} \bra{\downarrow} + k_2  \ket{\uparrow} \bra{\uparrow} \right) x \right],
\ee
fully removes the dependence of Eq.(\ref{EIT}) on the motional degrees of freedom
\be
\tilde{H}_{e}=\Delta \ket{e} \bra{e} +\frac{1}{2} \left( \Omega_1 \ket{e} \bra{\downarrow} + \Omega_2 \ket{e} \bra{\uparrow}   + H.c. \right).
\ee
and transfers it to $\tilde{H}_{m}$
\be\begin{split}
\tilde{H}_{m}=& \nu b^\dagger b + \nu \left( \eta_1^2 \ket{\downarrow} \bra{\downarrow} + \eta_2^2  \ket{\uparrow} \bra{\uparrow} \right) \\
&i\nu\left( \eta_1 \ket{\downarrow} \bra{\downarrow} + \eta_2  \ket{\uparrow} \bra{\uparrow} \right) \left( b^{\x}  - b\right).
\label{Htilm}
\end{split}\ee
In this picture, the crucial aspect surfaces that the interaction between motional and electronic degrees of freedom is modulated by the trap dynamics at a rate $\nu\eta_k$, which is independent of the laser intensity. It is worth noting that, even though the second term in Eq.(\ref{Htilm}) may seem to spoil the Raman resonance, it is negligible in the Lamb-Dicke limit and it can be reabsorbed by the appropriate adjustment of the laser frequencies $\w_1$ and $\w_2$.

For counter propagating beams, $\eta_1=-\eta_2=\eta$, the Hamiltonian becomes
\begin{eqnarray}
\tilde{H}_{e}&=&\Delta \ket{e} \bra{e} +\frac{\Omega}{2} \left( \ket{e} \bra{B} + H.c. \right), \label{eq:He}\\
\tilde{H}_{m}&=& \nu b^\dagger b + i\eta\nu\left( \ket{D} \bra{B} + H.c.\right)\left( b^\dagger - b\right).\label{eq:Hm}
\end{eqnarray}
The coupling structure between electronic and motional degrees of freedom has been considerably simplified with respect to Eq.(\ref{EIT}), as illustrated in Fig.\ref{Fig1}(b), since only a linear coupling remains between the electronic and motional degrees of freedom. It is important to note that this linear coupling is not the result of an approximation such as the Lamb-Dicke limit. In analogy to the conventional EIT proposal \cite{Morigi2000}, cooling is favored by adjusting the stark shift of $\ket B$ to $\nu$ as given by the eigenvalues of  $\tilde{H}_{e}$. This corresponds to  \cite{SM}
\be
\nu(\nu-\Delta)=\frac{\W^2}{4},
\label{MainResCond}
\ee
which is identical to the conventional EIT condition. Under this condition, resonant Rabi oscillations between states $\ket{D}\ket{n}$ and $\ket{B}\ket{n-1}$ take place, followed by decay from $\ket{B}\ket{n-1}$ to $\ket{D}\ket{n-1}$ mediated by the excited state. Nevertheless, these are not the original dark and bright states. Undoing the polaron transformation, $\ket{D}\ket{n}$ reveals itself as an in general entangled state $\ket{D,n}$ defined by
\be
\ket{D,n}\equiv\frac{1}{\W}\left(\W_1\ket\uparrow\ket{\eta_2,n}-\W_2\ket\downarrow\ket{\eta_1,n}\right)
\label{eq:dpol}
\ee
where we introduce the displaced Fock states $\ket{\alpha, n}=\exp(i\alpha b +i\alpha^* b\x)\ket{n}$ of the vibrational degrees of freedom of the ion. As displayed in Fig.\ref{Fig1}(c), these states are completely decoupled from the laser fields to all orders and thus we refer to them as {\em dark  polarons}. Although they are eigenstates of the electronic and laser Hamiltonians, they are not eigenstates of the trap Hamiltonian $H_m$ and therefore undergo an evolution $\ket{\alpha(t), n}$ with $\alpha(t)=\alpha e^{i\nu t}$. Under the laser cooling condition Eq.(\ref{MainResCond}), this implies that dark polaron $\ket{D,n}$ gradually becomes bright, i.e. it starts populating the orthogonal subspace
\be
\ket{B,n}\equiv\frac{1}{\W}\left(\W_2\ket\uparrow\ket{\eta_2,n}+\W_1\ket\downarrow\ket{\eta_1,n}\right),
\ee
which we refer to as {\em bright polarons}. In this picture, the laser-induced interaction between the electronic and vibrational degrees of freedom actually shows up as part of the trap dynamics.

{\em Numerical benchmark} --- We implement a numerical simulation of the full density matrix $\rho$ involving  both the electronic and motional degrees of freedom with the master equation
\be
\frac{d\rho}{dt}=-i\left[H',\rho\right]+\mathcal{L}^d(\rho)=\mathcal{L}(\rho).
\label{master}
\ee
The superoperator $\mathcal{L}^d$ is a Lindbladian for the two dissipative channels
\be
\mathcal{L}^d (\rho)= \sum_{i=\downarrow ,\uparrow} \gamma_{i} \left(2  \ket{i}\bra{e}\overline{\rho_{e,i}} \ket{e}\bra{i} -
\rho \ket{e}\bra{e}  - \ket{e}\bra{e}  \rho \right),
\label{diss}
\ee
where the spontaneous decay rates add up to the total rate $\gamma_{\downarrow}+\gamma_\uparrow=\Gamma$  and
\be
\overline{\rho_{e,i}} = \frac{1}{2}\int_{-1}^1 ds W(s)e^{ik_{e,i}xs} \rho e^{-ik_{e,i}xs}
\label{eq:rhoei}
\ee
accounts for the momentum transfer of $\hbar k_{e,i}$ in the event of a photon emission due to an electronic decay from level $\ket{e}$ to level $\ket{i}$. $W(s)=\frac 3 4 (1+s^2)$ is the angular distribution for a spontaneous emission of a dipole transition.

Note that Eq.\ref{eq:rhoei} is affected by the polaron transformation
\be
\tilde{\overline{\rho_{e,i}}} = \frac{1}{2}\int_{-1}^1 ds W(s)e^{i(sk_{e,i}-k_i)x} \rho e^{-i(sk_{e,i}-k_i)x}
\label{eq:rhoeipol}
\ee
Nevertheless, the leading order of both $\overline{\rho_{e,i}}$ and $\tilde{\overline{\rho_{e,i}}}$ with respect to the Lamb-Dicke expansion remains in both pictures and the same form of master equation should yield similar results in the Lamb-Dicke regime we will consider.

Indeed, as shown by Fig.\ref{Fig2}, both the original picture associated with $H'$ and the polaron picture associated with $\tilde{H}'$ coincide. In addition, as the Rabi frequency increases, they reproduce all of the regimes associated with EIT \cite{Cerrillo2018}. In the left panel, we observe faster cooling with increasing Rabi frequency, as predicted by Eq.(\ref{eq:Rad}). In the crossover regime (middle panel) the rate remains roughly constant over a large range of Rabi frequency values. In the right panel, an underdamped regime is shown, where the cooling rate decreases with increasing Rabi frequency and oscillations appear. In order to capture the main features of these dynamics, we turn to the derivation of a simple rate equation.

{\em Rate equation} --- From this polaron picture one can obtain a compact expression for the cooling rate. 
In standard implementations of EIT cooling the dominant decay channel is the excited–state linewidth 
$\Gamma$, which allows adiabatic elimination of the excited state. The resulting bright state relaxes 
directly into the dark state at a rate given by the inverse of its mean survival time 
 $\int_0^\infty c_{n}(t) dt$
\be
\gamma_B=8\nu^2\Gamma/\W^2.
\label{eq:gammaB}
\ee

This result leads to a Bloch–type set of equations for the red sideband in terms of the dark–polaron 
populations $p_n$, the bright polaron populations $p'_n$ and the real part of the bright-dark coherence $c_n$
\begin{eqnarray}
\frac{d}{dt}p_n&=& 2\sqrt{n}\eta \nu c_n + \gamma_B p'_n, \label{eq:BE}\\
\frac{d}{dt}c_n&=& \sqrt{n}\eta \nu \left( p'_{n-1}-p_n\right) -\frac{\gamma_B}{2} c_n , \nonumber\\
\frac{d}{dt}p'_{n-1}&=&  -2\sqrt{n}\eta \nu c_n - \gamma_B p'_{n-1}. \nonumber
\end{eqnarray}
An adiabatic elimination of $p'_n$ and $c_n$ yields an average decay rate

\begin{figure}
    \includegraphics[width=\columnwidth]{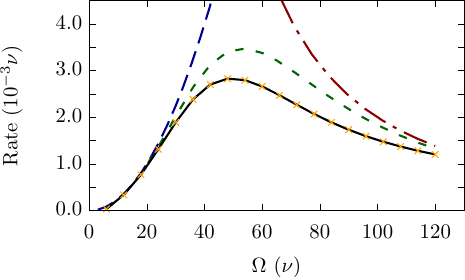}
    \caption{ 
    Cooling rate as a function of Rabi frequency $\W$ for the full master equation Eq.(\ref{master}) in the original picture (solid black line), in the polaron picture (orange crosses), for rate equation Eq.(\ref{RE}) (dashed green line), for low intensity approximation Eq.(\ref{eq:Rad}) (long-dashed blue line) and for high intensity approximation Eq.(\ref{eq:gammaB}) (dash-dotted red line). Parameters: $\Gamma=10\nu$, $\eta_1=-\eta_2=1/\sqrt{2}$.}
    \label{Rates}
\end{figure}

(NOTE: The expression with the real part of the inverse of the effective Hamiltonian directly provides A-=etaW/Gamma. at the same time, it can be expressed as inverse of eigenvalues of Heff, with real and imaginary parts. These two procedures need to be reconciled)
\be
A_n^-=\cfrac{2n\eta^2\nu^2}{\cfrac{\gamma_B}{2}+\cfrac{2n\eta^2\nu^2}{\gamma_B}},
\label{Am}
\ee
A similar analysis for the blue sideband provides an average heating rate
\be
A_n^+=\cfrac{2n\eta^2\nu^2}{\cfrac{\gamma_B}{2}+\cfrac{8\nu}{\gamma_B}+\cfrac{2n\eta^2\nu^2}{\gamma_B}}.
\label{Ap}
\ee
These two rates provide us with a rate equation for $p_n$ 
\be
\frac{d}{dt}p_n=A_{n+1}^- p_{n+1} + A_{n}^+ p_{n-1} - \left(A_{n+1}^++A_n^-\right)p_n.
\label{RE}
\ee
Although the form of this rate equation is analogous to that of conventional treatments \cite{Cirac1992}, it is based in a perturbative expansion with respect to $\eta\nu$ rather than $\eta \W$ and correctly captures the cooling dynamics for arbitrary values $\W$.
Its overall cooling rate prediction is compared with the full numerical simulation of Fig.\ref{Fig2}  by exponential fittings of $\ave{n}$. Results are shown for a large range of laser intensities in Fig.\ref{Rates}. Eq.(\ref{RE}) is found to approximate the cooling process well in the whole range, in particular by reproducing the turnover between the limiting regimes of small and large Rabi frequency.

{\em Low laser intensity} ---  For small $\W$, $\gamma_B\gg\eta\nu$ and the last term in the denominators of Eq.(\ref{Ap},\ref{Am}) can be neglected. In this limit the cooling rates take the usual form $A^\pm_n=nA^\pm$ such that a closed equation for the phonon population may be derived \cite{Cirac1992} 
\be
\frac{d}{dt}\ave{n}=-(A^--A^+)\ave{n}+A^+.
\label{eq:nad}
\ee
Under the regression theorem, rates $A^\pm$ can be interpreted as the real part of the Fourier transform of
\be
C(t) =\ave{\sigma_{\eta}(t)\sigma_{\eta}},\label{CorrFun}
\ee
the correlation function of $\sigma_\eta =\eta \nu\left( \ket{B} \bra{D} + H.c.\right)$, which is the electronic part of the interaction term of the Hamiltonian $\tilde{H}'$  (see Eq.\ref{eq:Hm}), averaged over the electronic steady state $\ket{D}$. Even though this derivation is performed in the dark polaron picture, the resulting rates fully coincide with those in the original picture \cite{Morigi2000} and so $A_-$ matches Eq.\ref{eq:Rad} and predicts a quadratic increase with the laser intensity as shown in Fig.\ref{Rates}.

{\em Large laser intensity} ---  In the limit of large $\W$, $\gamma_B$ monotonously decreases such that  the last term in the denominators of Eq.(\ref{Am},\ref{Ap}) becomes dominant, and $A_n^-\simeq\gamma_B$ for all $n$. As shown in Fig.\ref{Rates}, the cooling rate is therefore upper bounded by $\gamma_B$.

{\em Conclusions} ---
In order to describe the reason behind observed upper bounds in the cooling rate of trapped ions with EIT, we develop a theory based on the concept of dark and bright polarons. In this picture, the interaction term between trap and polaron degrees of freedom is not proportional to the laser intensity, facilitating analysis of its effect for all possible values. In addition, no high-order Lamb-Dicke terms exist that spoil the laser cooling picture for high-lying Fock levels. We propose a master equation in that picture and benchmark with the common master equation, finding perfect agreement between both. Further, a rate equation is proposed that bridges cooling dynamics between the large and small intensity limits. In the weak intensity limit, it coincides with existing knowledge, while it is found that in the large intensity limit cooling is dominated by the slow decay of the bright polarons. This confirms previous numerical and experimental observations and opens up a much needed new tool for the design of fast cooling and control methods for trapped ions, even beyond the Lamb-Dicke limit or for initial temperatures well above the Doppler limit.

\begin{acknowledgments}
The authors acknowledge support from grant CNS2023-144994 funded by MICIU/AEI/10.13039/201100011033 and by ``ERDF/EU''. J.C. additionally acknowledges support from European Union project C-QuENS (Grant No. 101135359).
\end{acknowledgments}

\end{document}